\documentclass[fleqn,12pt]{article}
\usepackage{amsmath,amssymb,epsfig,times,graphics}
\usepackage{amssymb}
\usepackage{amsmath}
\usepackage{CJK}
\usepackage{graphics}
\usepackage{graphicx}
\usepackage{dcolumn}
\usepackage{bm}
\usepackage{subfigure}
\usepackage{lineno}
\usepackage{amsthm}
\usepackage{epstopdf}
\begin{document}

\title{\bf The phenotypic equilibrium of cancer cells: From average-level stability to path-wise convergence}

\date{}
\maketitle

\author{Yuanling Niu$^{2}$, Yue Wang$^3$, Da Zhou$^{1,*}$}

\begin{enumerate}
  \item School of Mathematical Sciences, Xiamen University,
Xiamen 361005, P.R. China\\ ($*$Corresponding Author, zhouda@xmu.edu.cn)
  \item School of Mathematics and Statistics, Central South University, Changsha 410083, P.R. China
  \item Department of Applied Mathematics, University of Washington,
Seattle, WA 98195, USA
\end{enumerate}

\begin{abstract}
The phenotypic equilibrium, i.e. heterogeneous population of cancer cells tending to a fixed
equilibrium of phenotypic proportions, has received much attention in cancer biology
very recently. In previous literature, some theoretical models were used to predict the
experimental phenomena of the phenotypic equilibrium, which were often explained by
different concepts of stabilities of the models. Here we present a stochastic multi-phenotype
branching model by integrating conventional cellular hierarchy with phenotypic plasticity
mechanisms of cancer cells. Based on our model, it is shown that:
\emph{(i)} our model can serve as a framework to unify the previous models for
the phenotypic equilibrium, and then harmonizes the different kinds of average-level
stabilities proposed in these models; and \emph{(ii)} path-wise convergence of our model
provides a deeper understanding to the phenotypic equilibrium from stochastic point of view.
That is, the emergence of the phenotypic equilibrium is rooted in the stochastic nature
of (almost) every sample path, the average-level stability
just follows from it by averaging stochastic samples.

\end{abstract}


\section{Introduction}
\label{intro}

Stability is ubiquitous in biology, ranging from
physicochemical homeostasis in cellular microenvironments to ecological
constancy and resilience
\cite{cannon1929organization,gorshkov1994physical,justus2008ecological}.
It is noteworthy that not only can the stability phenomenon arise in normal living systems,
but it can also happen in abnormal organisms such as cancer.
As a large family of diseases with abnormal cell growth,
cancer is generally acknowledged to be the malignant progression
along with a series of stability-breaking changes (\emph{e.g.} genomic instability)
within the normal organisms \cite{hanahan2011hallmarks}.
However, some recent researches reveal the other side of cancer.
An interesting \emph{phenotypic equilibrium}
was reported in some cancers
\cite{chaffer2011normal,gupta2011stochastic,yang2012dynamic}.
That is, the population composed of different cancer cells will tend to a fixed
equilibrium of phenotypic proportions over time regardless of initial states
(Fig. 1). These findings provided new insights to the research of cancer heterogeneity.
\begin{figure}
\begin{center}
\includegraphics[width=1.2\textwidth]{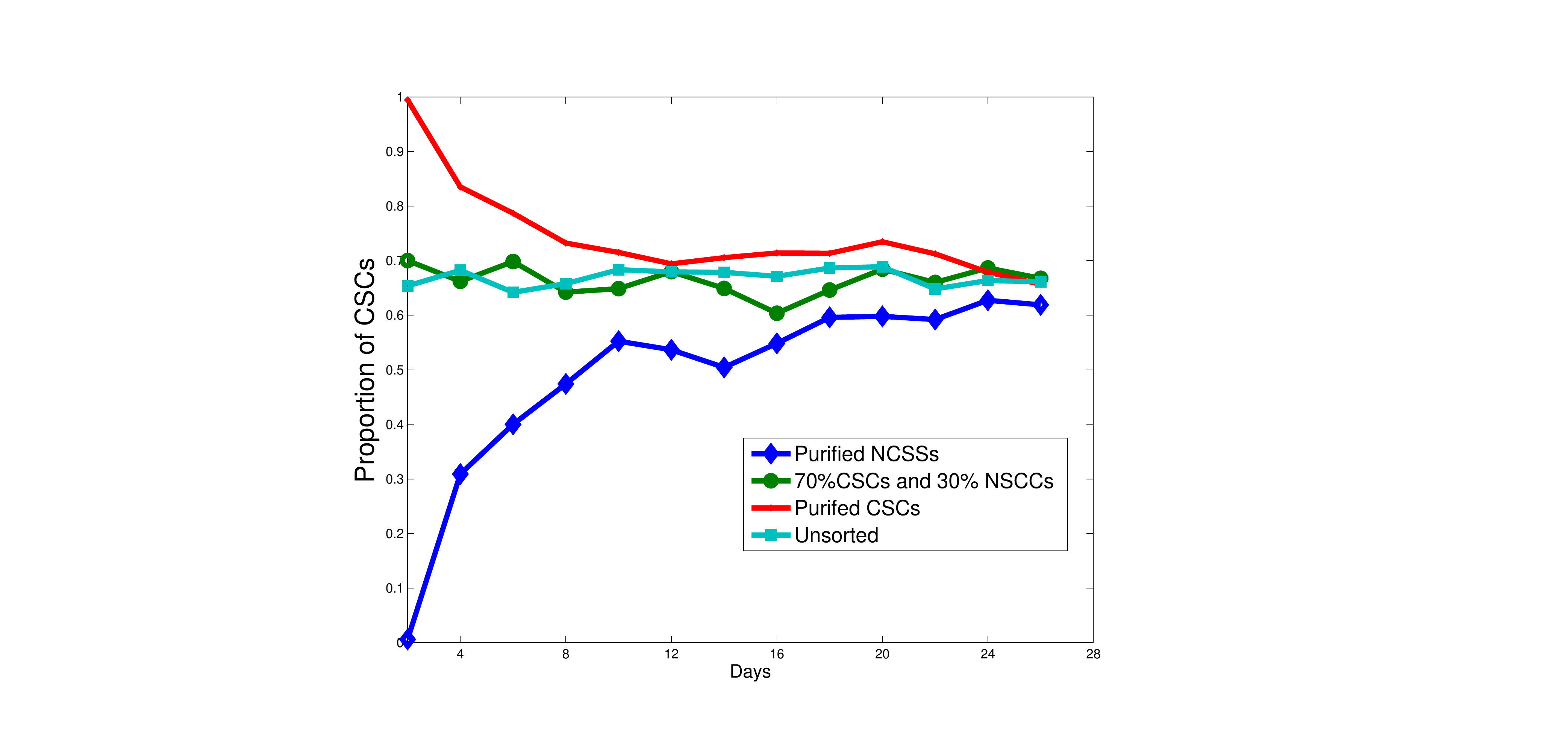}
\caption{The phenotypic equilibrium of cancer cells. The figure is generated from the
data (SW620 colon cancer cell line) in \cite{yang2012dynamic}. In this experiment, two cellular phenotypes
were identified: cancer stem cells (CSCs) and non-stem cancer cells (NSCCs). It is shown that
no matter where the initial state is (as four different cases shown in the figure),
the CSCs proportion will converge to a fixed proportion as time passes.
The same is true for NSCCs proportion.
This phenomenon is termed \emph{phenotypic equilibrium} \cite{gupta2011stochastic}.}
\end{center}
\end{figure}

The experimental works also stimulate theoreticians to put forward reasonable models for interpreting
the phenotypic equilibrium \cite{gupta2011stochastic,zapperi2012cancer,
dos2013possible,dos2013noise,zhou2013population,wang2014dynamics,zhou2014multi,zhou2014nonequilibrium}.
In particular, it was reported that the intrinsic interconversion between different cellular phenotypes,
also called \emph{phenotypic plasticity} \cite{french2012complex,meacham2013tumour}, could play a crucial role in stabilizing the
mixture of phenotypic proportions in cancer. As a pioneering work, Gupta \emph{et al} proposed a discrete-time Markov chain model to
describe the phenotypic transitions in breast cancer cell lines \cite{gupta2011stochastic}. In their model, three phenotypes were
identified: stem-like cells (S), basal cells (B) and luminal cells (L). The phenotypic transitions among them can be captured by
the transition probability matrix as follows:
\begin{linenomath*}
\begin{equation}
P=\left(\begin{array}{ccc}
1-P_{S\rightarrow B}-P_{S\rightarrow L} & P_{S\rightarrow B} & P_{S\rightarrow L}  \\
P_{B\rightarrow S} & 1-P_{B\rightarrow S}-P_{B\rightarrow L} & P_{B\rightarrow L}  \\
P_{L\rightarrow S} & P_{L\rightarrow B} & 1-P_{L\rightarrow S}-P_{L\rightarrow B} \\
\end{array}\right),
\label{Matrix1}
\end{equation}
\end{linenomath*}
where $P_{i\rightarrow j}$ represents the probability of the transition from phenotype $i$ to $j$.
According to the limiting theory of discrete-time finite-state Markov chain, there exists unique equilibrium
distribution $\vec{\pi}=(\pi_S, \pi_B, \pi_L)$ such that $\vec{\pi}=\vec{\pi}P$, provided $P$ is irreducible
and aperiodic \cite{seneta1981non}. The Markov chain will converge to $\vec{\pi}$ regardless of where it begins.
By fitting the Markov chain model to their experimental data, the equilibrium proportions of
stem-like, basal and luminal cells were predicted by the equilibrium distribution $\pi_S, \pi_B, \pi_L$ respectively.

Even though the Markov chain model fitted the experimental results in breast cancer cell lines very well,
Zapperi and La Porta \cite{zapperi2012cancer} questioned the validity of the phenotypic transitions
and gave an alternative explanation to the phenotypic equilibrium, which was based on
the conventional cancer stem cell (CSC) model with imperfect CSC biomarkers.
Moreover, Liu \emph{et al} showed that the negative feedback mechanisms of
non-linear growth kinetics of cancer cells can also control the balance between
different cellular phenotypes \cite{liu2013nonlinear}.
These works suggested that the phenotypic plasticity may not be the only explanation to the phenotypic
equilibrium. To further reveal the mechanisms giving rise to
the phenotypic equilibrium, it is more convincing to study the models
integrating the phenotypic plasticity with the other conventional cellular processes
of cancer. Motivated by this, a series of works discussed the phenotypic equilibrium
by establishing the models coordinating with both
hierarchical cancer stem cell paradigm and phenotypic plasticity
\cite{dos2013possible,dos2013noise,zhou2013population,wang2014dynamics,zhou2014multi,zhou2014nonequilibrium}.
In these works, the phenotypic equilibria were intimately related to the stable steady-state
behavior of the corresponding ordinary differential equations (ODEs) models. In other
words, if one can model the dynamics of the phenotypic proportions as the following
system of ODEs
\begin{linenomath*}
$$\frac{d\vec{x}}{dt}=\vec{F}(\vec{x}),$$
\end{linenomath*}
the unique stable fixed point $\vec{x}^*$ (if exists) corresponds to the equilibrium
proportions.

The aforementioned works have showed that the phenotypic equilibrium can be explained by
different concepts of stabilities in different models. Thus a natural
question is whether there exists a unified framework to
harmonize the equilibrium distribution of the Markov chain model and the stable
steady-state behavior of the ODEs model. In this study, we try to address this issue
by establishing a multi-phenotype branching (MPB) model \cite{athreya1972branching}.
On one hand, our model integrates the phenotypic plasticity with
the cellular processes (such as cell divisions) that have extensively been studied in cancer biology.
On the other hand, the model is stochastic and closer to the reality with finite population size
\cite{dingli2007symmetric,antal2011exact}. Based on this model, it is shown that the ODEs
model can be derived by taking the expectation of our model.
More specifically, the ODEs model is just the \emph{proportion equation} of
the MPB model. Besides, the Markov chain model is also shown to be
closely related to our model. That is, the Kolmogorov forward equation of the continuous-time Markov chain model
is a special case of the proportion equation provided that the division rates of stem-like, basal and luminal
cells are the same. Interestingly, ``same doubling time'' of the three phenotypes was just observed
in Gupta \emph{et al}'s  experiment when they used the Markov chain model to explain the phenotypic equilibrium
\cite{gupta2011stochastic}, which is in line with our theoretical prediction.
Moreover, our result also shows that one should be more cautious about
the application of the Markov chain in modeling cell-state dynamics in larger time scales, since the Markov chain
model takes no account of different capabilities of divisions by cancer stem cells and non-stem cancer cells.

More importantly, by showing \emph{almost sure convergence} of
the MPB model, the stationarity of the Markov chain model and the stability of the ODEs model
can be unified as the average-level stability of our model. Note that
the almost sure convergence indicates the path-wise stability of stochastic samples,
providing a more profound explanation to the phenotypic equilibrium. In other words,
the phenotypic equilibrium is actually rooted in the stochastic nature of (almost) every path sample;
the average-level stability just follows from it by averaging all the stochastic samples.
Furthermore, it is also shown that, not only can the model with phenotypic plasticity give rise to the
path-wise convergence, but the conventional cancer stem cell model without phenotypic plasticity
can also lead to the convergence under certain conditions.
This echoes the works \cite{zapperi2012cancer, liu2013nonlinear}
that the phenotypic plasticity is not the only explanation to the phenotypic equilibrium.

The paper is organized as follows. The model is presented in Section 2.
Main results are shown in Section 3. Conclusions are in Section 4.

\section{Model}
\label{Model}

\subsection{Assumptions}

In this section we give the assumptions of our model. Consider a population composed of different cancer cell phenotypes.
For pure theoretical investigations, the number of the phenotypes can be any $n$ in general \cite{zhou2014multi,jiang2014cell}.
However, to better illustrate our theoretical results on the basis of more concrete biological background, enlightened by
\cite{gupta2011stochastic}, we here focus on the specific model consisting of three phenotypes:
stem-like cells (S), basal cells (B) and luminal cells (L). The main assumptions are listed as follow:\\
\emph{1}. Stem-like cells can perform three types of divisions: symmetric division, symmetric differentiation and asymmetric division
\cite{morrison2006asymmetric, dalerba2007cancer,todaro2010colon}.
That is, a stem-like cell can divide into two identical stem-like cells (symmetric division) or two identical differentiated cancer cells
(symmetric differentiation; it can also divide into a stem-like cell and a differentiated cancer cell (asymmetric division).
\begin{itemize}
  \item symmetric division: S $\overset{\alpha_{S}P_1}{\longrightarrow}$ S+S;
  \item symmetric differentiation: S $\overset{\alpha_{S}P_2}{\longrightarrow}$ B+B or S $\overset{\alpha_{S}P_3}{\longrightarrow}$ L+L;
  \item asymmetric division: S $\overset{\alpha_{S}P_4}{\longrightarrow}$ S+B or S $\overset{\alpha_{S}P_5}{\longrightarrow}$ S+L.
\end{itemize}
$\alpha_S$ is the division rate (or termed synthesis rate \cite{liu2013nonlinear}),
with the meaning that a stem-like cell will wait an exponential time
with expectation $\alpha_S$ and then perform one particular type of division with probability $P_i$ (note that $\sum_{i=1}^5P_i=1$).
Suppose the waiting time and the division strategy are independent to each other,
then the product of $\alpha_S$ and $P_i$ governs the reaction rate of the corresponding division type. \\
\emph{2}. For non-stem cancer cells, \emph{i.e.} basal or luminal cells, we assume that not only can they undergo symmetric divisions
with limited times, but they can also perform phenotypic conversions. To illustrate this,
let us take B phenotype as an example. Suppose a
newly-born B cell can divide at most $m$ times. If we denote $B_i$ as the B cell that has already divided $i$ times, then we have
the following hierarchical structure:
\begin{itemize}
  \item $\textrm{B}_0$ $\overset{\alpha_{B}}{\longrightarrow}$ $\textrm{B}_1$+$\textrm{B}_1$;
  \item ...
  \item $\textrm{B}_{m-1}$ $\overset{\alpha_{B}}{\longrightarrow}$ $\textrm{B}_m$+$\textrm{B}_m$;
  \item $\textrm{B}_m$ $\overset{\alpha_{B_m}}{\longrightarrow}$ $\emptyset$.
\end{itemize}
$\alpha_{B}$ is the division rate, and $\alpha_{B_m}$ is the death rate of $\textrm{B}_m$.
Moreover, assume that a B cell can convert into an S cell (termed \emph{de-differentiation} \cite{marjanovic2013cell})
by phenotypic plasticity. Let the dedifferentiation rate of $\textrm{B}_i$ be $\beta_{B_i}$, then we have
\begin{itemize}
  \item $\textrm{B}_0$ $\overset{\beta_{B_0}}{\longrightarrow}$ S;
  \item ...
  \item $\textrm{B}_m$ $\overset{\beta_{B_m}}{\longrightarrow}$ S.
\end{itemize}
For simplicity, it is often assumed that $\beta_{B_0}=...=\beta_{B_m}$ \cite{wang2014dynamics},
denoted as $\beta_{B}$ for short. Meanwhile, note that a B cell can also convert into an L cell \cite{gupta2011stochastic}.
Since the biological mechanisms of the phenotypic conversions between different non-stem cancer cells are still
poorly understood, for simplicity it is assumed that the phenotypic transitions between B and L can only happen
in the same hierarchical level. That is, supposing that a newly-born L cell can also
divide at most $m$ times, $L_i$ is the L cell that has already divided $i$ times, then we have
\begin{itemize}
  \item $\textrm{B}_i$ $\overset{\gamma_{B}}{\longrightarrow}$ $\textrm{L}_i$;
\end{itemize}
$\gamma_{B}$ is the transition rate. In fact, this assumption implies
$\textrm{B}$ ${\longrightarrow}$ $\textrm{L}$ with constant rate $\gamma_B$ overall,
which is in line with the assumption in \cite{gupta2011stochastic}.
For luminal cells, similarly, their cellular
processes are shown as follows:
\begin{itemize}
  \item $\textrm{L}_i$ $\overset{\alpha_L}{\longrightarrow}$ $\textrm{L}_{i+1}$+$\textrm{L}_{i+1}$~~~($0\leq i\leq m-1$);
  \item $\textrm{L}_m$ $\overset{\alpha_{L_m}}{\longrightarrow}$ $\emptyset$.
  \item $\textrm{L}_i$ $\overset{\beta_{L}}{\longrightarrow}$ S~~~($0\leq i\leq m$);
  \item $\textrm{L}_i$ $\overset{\gamma_{L}}{\longrightarrow}$ $\textrm{B}_i$~~~($0\leq i\leq m$).
\end{itemize}

\subsection{Multi-phenotypic branching (MPB) model}

Based on the cellular processes listed in the last subsection, we can model this cellular system
as a continuous-time Markov process in the discrete state space of cell numbers (Chapter 11 in \cite{beard2008chemical}).
If we let $X_1$ be the cell number of S phenotype,
$\vec{X}_2=(X^{(0)}_2, X^{(1)}_2,...,X^{(m)}_2)^T$ be the vector representing the cell numbers of $\textrm{B}_i$ cells,
and $\vec{X}_3=(X^{(0)}_3, X^{(1)}_3,...,X^{(m)}_3)^T$ be the vector representing the cell numbers of $\textrm{L}_i$ cells,
then the dynamics of $\vec{X}=(X_1, \vec{X}_2,\vec{X}_3)^T$ can be modeled as a multi-phenotype branching process
\cite{athreya1972branching}.
If we define $\textrm{Pr}(\vec{x};t)$ be the probability of $\vec{X}=\vec{x}$ at time $t$,
according to the theory of \emph{Chemical Master Equation} (CME), the rate of change of $\textrm{Pr}(\vec{x};t)$
 is equal to the transitions into $\vec{x}$ minus the transitions out of it, \emph{i.e.}
\begin{linenomath*}
\begin{equation}
\frac{d\textrm{Pr}(\vec{x};t)}{dt}=\sum_{\vec{x}'\neq \vec{x}}T_{\vec{x}'\rightarrow
\vec{x}}\textrm{Pr}(\vec{x}';t)-\sum_{\vec{x}'\neq \vec{x}}T_{\vec{x}\rightarrow \vec{x}'}\textrm{Pr}(\vec{x};t),
\label{CME}
\end{equation}
\end{linenomath*}
where $T_{\vec{x}'\rightarrow \vec{x}}$ is the transition rate from $\vec{x}'$ to $\vec{x}$ and
$T_{\vec{x}\rightarrow \vec{x}'}$ is the transition rate from $\vec{x}$ to $\vec{x}'$
(see \ref{appendix1} for more details).

In next section we will show that the ODEs model and the Markov chain model can be
derived from our model. For convenience we term our multi-phenotype branching model
the\emph{ MPB model}.

\section{Results}

\subsection{Deterministic equations derived from the MPB model}
\label{section3.1}

To relate our MPB model to the ODEs model, we consider the mean dynamics of the MPB model
by averaging all the stochastic samples of it.

Let $\langle{\vec{X}}\rangle$ be the expectation of $\vec{X}$, that is, for each component
we define
\begin{linenomath*}
$$\langle X_i \rangle:=\sum_{\vec{x}}x_i\textrm{Pr}(\vec{x}; t).$$
\end{linenomath*}
We multiply $x_i$ on the both sides of Eq. (\ref{CME}), and then calculate the summation over all $\vec{x}$
\begin{linenomath*}
\begin{equation*}
\sum_{\vec{x}}x_i\frac{d\textrm{Pr}(\vec{x};t)}{dt}=\sum_{\vec{x}}x_i\left(\sum_{\vec{x}'\neq \vec{x}}T_{\vec{x}'\rightarrow
\vec{x}}\textrm{Pr}(\vec{x}';t)-\sum_{\vec{x}'\neq \vec{x}}T_{\vec{x}\rightarrow \vec{x}'}\textrm{Pr}(\vec{x};t)\right).
\end{equation*}
\end{linenomath*}
For S cells:
\begin{linenomath*}
\begin{equation}
\frac{d\langle X_1\rangle}{dt}=\alpha_S\left(P_1-P_2-P_3\right)\langle X_1\rangle+\beta_{B}\sum_{i=0}^{m}X^{(i)}_2
+\beta_{L}\sum_{i=0}^{m}X^{(i)}_3.
\end{equation}
\end{linenomath*}
For B cells:
\begin{linenomath*}
\begin{equation}
\begin{cases}
\frac{d\langle X^{(0)}_2\rangle}{dt}=\alpha_S\left(2P_2+P_4\right)\langle X_1\rangle-
\left(\alpha_{B}+\beta_{B}+\gamma_{B}\right)\langle X^{(0)}_2\rangle
+\gamma_{L}X^{(0)}_3;\\
\frac{d\langle X^{(i)}_2\rangle}{dt}=2\alpha_{B}\langle X^{(i-1)}_2\rangle-
\left(\alpha_{B}+\beta_{B}+\gamma_{B}\right)\langle X^{(i)}_2\rangle
+\gamma_{L}X^{(i)}_3~~~~(1\leq i\leq m-1);\\
\frac{d\langle X^{(m)}_2\rangle}{dt}=2\alpha_{B}\langle X^{(m-1)}_2\rangle-
\left(\alpha_{B_m}+\beta_{B}+\gamma_{B}\right)\langle X^{(m)}_2\rangle
+\gamma_{L}X^{(m)}_3.
\end{cases}
\end{equation}
\end{linenomath*}
For L cells:
\begin{linenomath*}
\begin{equation}
\begin{cases}
\frac{d\langle X^{(0)}_3\rangle}{dt}=\alpha_S\left(2P_3+P_5\right)\langle X_1\rangle-
\left(\alpha_{L}+\beta_{L}+\gamma_{L}\right)\langle X^{(0)}_3\rangle
+\gamma_{B}X^{(0)}_2;\\
\frac{d\langle X^{(i)}_3\rangle}{dt}=2\alpha_{L}\langle X^{(i-1)}_2\rangle-
\left(\alpha_{L}+\beta_{L}+\gamma_{L}\right)\langle X^{(i)}_3\rangle
+\gamma_{B}X^{(i)}_2~~~~(1\leq i\leq m-1);\\
\frac{d\langle X^{(m)}_3\rangle}{dt}=2\alpha_{L}\langle X^{(m-1)}_2\rangle-
\left(\alpha_{L_m}+\beta_{L}+\gamma_{L}\right)\langle X^{(m)}_3\rangle
+\gamma_{B}X^{(m)}_2.
\end{cases}
\end{equation}
\end{linenomath*}
Then it is not difficult to see that the dynamics of $\langle\vec{X}\rangle$ can be
captured by a system of linear ODEs,
\begin{linenomath*}
\begin{equation}
\frac{d \langle\vec{X}\rangle}{d t}=G\langle\vec{X}\rangle,
\label{ODE1}
\end{equation}
\end{linenomath*}
where
\begin{linenomath*}
\begin{equation}
G=[g_{ij}]_{(2m+3)\times(2m+3)}=\left(\begin{smallmatrix}
\alpha_S\left(P_1-P_2-P_3\right) & \beta_{B} & \cdots & \beta_{B} & \beta_{L} & \cdots & \beta_{L} \\
\alpha_S\left(2P_2+P_4\right) & -\left(\alpha_{B}+\beta_{B}+\gamma_{B}\right) & 0 & \cdots & \gamma_{L} & \cdots & 0  \\
0 & 2\alpha_{B} & -\left(\alpha_{B}+\beta_{B}+\gamma_{B}\right) & 0 & \cdots & \cdots & 0 \\
\cdots & \cdots & \cdots & \cdots & \cdots & \cdots & \cdots\\
\end{smallmatrix}\right).
\label{Matrix2}
\end{equation}
\end{linenomath*}

Furthermore, it should be noted that, Eq. (\ref{ODE1}) describes the cell number dynamics of each phenotype at each hierarchical level.
If we denote $X_2=\sum_{i=0}^m X^{(i)}_2$ and $X_3=\sum_{i=0}^m X^{(i)}_3$ as the total cell numbers of B and L phenotypes
respectively, then it is often the dynamics of $\vec{X}^*=(X_1, X_2, X_3)^T$ that interests people. That is,
\begin{linenomath*}
\begin{small}
\begin{equation}
\begin{cases}
\frac{d\langle X_1\rangle}{dt}=\alpha_S\left(P_1-P_2-P_3\right)\langle X_1\rangle+\beta_{B}\langle X_2\rangle+\beta_{L}\langle X_3\rangle;\\
\frac{d\langle X_2\rangle}{dt}=\alpha_S\left(2P_2+P_4\right)\langle X_1\rangle+(\alpha_B-\beta_B-\gamma_B)\langle X_2\rangle
+\gamma_{L}\langle X_3\rangle-(\alpha_B+\alpha_{B_m})\langle X^{(m)}_2\rangle;\\
\frac{d\langle X_3\rangle}{dt}=\alpha_S\left(2P_3+P_5\right)\langle X_1\rangle+\gamma_{B}\langle X_2\rangle
+(\alpha_L-\beta_L-\gamma_L)\langle X_3\rangle-(\alpha_L+\alpha_{L_m})\langle X^{(m)}_3\rangle.\\
\end{cases}
\label{ODEx}
\end{equation}
\end{small}
\end{linenomath*}
We can see that Eq. (\ref{ODEx}) is not linear of $\langle \vec{X}^* \rangle$, which also depends on $\langle X^{(m)}_2\rangle$ and
$\langle X^{(m)}_3\rangle$ separately. Technically this is due to the limited capability of divisions of B and L phenotypes.
In the limit of $m$, or when $m$ is relatively large in comparison to observational time scales
(\emph{e.g.} $t\lessapprox m$), Eq. (\ref{ODEx}) can approximately be
expressed as a linear system of $\langle \vec{X}^* \rangle$:
\begin{linenomath*}
\begin{equation}
\frac{d \langle\vec{X}^*\rangle}{d t}\approx G^*\langle\vec{X}^*\rangle,
\label{ODExx}
\end{equation}
\end{linenomath*}
where
\begin{linenomath*}
\begin{equation}
G^*=[g^*_{ij}]_{3\times3}=\left(\begin{smallmatrix}
\alpha_S\left(P_1-P_2-P_3\right) & \beta_{B} & \beta_{L}\\
\alpha_S\left(2P_2+P_4\right) & \alpha_B-\beta_B-\gamma_B & \gamma_{L}\\
\alpha_S\left(2P_3+P_5\right) & \gamma_{B} & \alpha_L-\beta_L-\gamma_L
\end{smallmatrix}\right).
\label{Matrixx}
\end{equation}
\end{linenomath*}
In this way the model reduces to the three-phenotypic model investigated in \cite{zhou2013population}.
However, Eq. (\ref{ODEx}) should be adopted for describing larger time scales (\emph{e.g.} $t\gg m$).
Note that it is inconvenient to analyze Eq. (\ref{ODEx}) directly,
we will show later that analyzing Eq. (\ref{ODE1}) is quite helpful for
the understanding of Eq. (\ref{ODEx}), especially in the study of the phenotypic equilibrium.

\subsection{Proportion equation: Bridging the MPB model and the ODEs model}

Since Eq. (\ref{ODE1}) describes the dynamics of the absolute numbers of different cellular
phenotypes, we term it the \emph{number equation}. However, to investigate the phenotypic
equilibrium, we are more concerned about the dynamics of the relative numbers (\emph{i.e.} proportions)
of different cellular phenotypes. Let $\vec{p}$ be the vector representing the proportions of different
cellular phenotypes. By replacing $\langle \vec{X}\rangle$ in Eq. (\ref{ODE1}) with $\vec{p}$ ,
we have the equation governing the phenotypic proportions as follows (see \ref{appendix2})
\begin{linenomath*}
\begin{equation}
\frac{d \vec{p}}{d t}=G\vec{p}-\vec{p} e^TG\vec{p},
\label{ODE2}
\end{equation}
\end{linenomath*}
where $e=(1,...,1)^T$. We term Eq. (\ref{ODE2}) the \emph{proportion equation}. It is noteworthy that
the stable steady-state behavior of Eq. (\ref{ODE2}) just corresponds to the phenotypic equilibrium
investigated in \cite{wang2014dynamics, zhou2014multi}.
The proportion equation thus connects the MPB model and the ODEs model in previous literature,
implying that the ODEs model can be seen as the average-level counterpart
of the stochastic MPB model. To show the stability of Eq. (\ref{ODE2}),
we have the following theorem (see \ref{appendix2+} for the proof):

\newtheorem{theorem}{Theorem}
\begin{theorem}
There exists unique positive stable fixed point $\vec{\mu}$ in Eq. (\ref{ODE2})
provided that $G$ is irreducible
\footnote{Strictly speaking, for completing the theorem
it is necessary to add a small perturbation to the initial state in rare cases,
see \ref{appendix2+}}.
\label{Thm1}
\end{theorem}

Theorem \ref{Thm1} shows that the deterministic population dynamics of cancer cells
will tend to an equilibrium mixture of phenotypic proportions
as time passes. Besides, let $\vec{p}^*$ be the proportion vector of $\vec{X}^*$,
\emph{i.e.}
 \begin{linenomath*}
$$\vec{p}^*=(p^*_1, p^*_2, p^*_3)=(p_1, \sum_{i=0}^m p^{(i)}_2, \sum_{i=0}^m p^{(i)}_3).$$
\end{linenomath*}
Given $\lim_{t\rightarrow \infty}\vec{p}=\vec{\mu}$ (Theorem \ref{Thm1}),
\begin{linenomath*}
$$\lim_{t\rightarrow \infty}\vec{p}^*=\lim_{t\rightarrow \infty}(p_1, \sum_{i=0}^m p^{(i)}_2, \sum_{i=0}^m p^{(i)}_3)=
(\mu_1, \sum_{i=0}^m \mu^{(i)}_2, \sum_{i=0}^m \mu^{(i)}_3)=\vec{\mu}^*.$$
\end{linenomath*}
Thus we have the following result for $\vec{p}^*$:
\newtheorem{corollary}{Corollary}
\begin{corollary}
Under the same condition in Theorem \ref{Thm1}, $\vec{p}^*$ will tend to a fixed positive vector $\vec{\mu}^*$ as $t\rightarrow\infty$.
\label{cor1}
\end{corollary}

Corollary \ref{cor1} indicates the phenotypic equilibrium of the three-phenotypic model in Eq. (\ref{ODEx}).
Moreover, it should be pointed out that, the results in Theorem \ref{Thm1} and Corollary \ref{cor1}
can be seen as the average-level stabilities following from the the path-wise convergence of the MPB model,
which will be discussed in Sec. \ref{sectionIII}.

\subsection{The Markov chain model as a special case of the proportion equation}

Note that the Markov chain model Eq. (\ref{Matrix1}) is discrete-time and the MPB model is continuous-time;
to compare the two models in the same time scale, we turn our attention from discrete-time Markov chain
to continuous-time Markov chain. Consider the standard model of continuous-time Markov chain. That is, let
$P_i(t)$ be the probability of the Markov chain being in state $i$ at time $t$,
its dynamics can be captured by the Kolmogorov forward equation:
\begin{linenomath*}
\begin{equation}
\frac{d \vec{P}(t)}{d t}=Q^T\vec{P}(t),
\label{MC}
\end{equation}
\end{linenomath*}
where $Q$-matrix $[q_{ij}]_{3\times 3}$ satisfying
\begin{linenomath*}
\begin{align}
q_{ij}\geq 0 ~~ \forall i\neq j,
\label{Q1}
\end{align}
\end{linenomath*}
\begin{linenomath*}
\begin{align}
q_{ii}=-\sum_{j:j\neq i}q_{ij}.
\label{Q2}
\end{align}
\end{linenomath*}

We now discuss the relation between $\vec{P}(t)$ and $\vec{p}^*$.
By replacing $\langle \vec{X}^*\rangle$ in Eq. (\ref{ODExx}) with $\vec{p}^*$,
we obtain the proportion equation governing $\vec{p}^*$
\footnote{The derivation of Eq. (\ref{ODE*}) is similar to that of Eq. (\ref{ODE2}), see \ref{appendix2}.}
\begin{linenomath*}
\begin{equation}
\frac{d \vec{p}^*}{d t}=G^*\vec{p}^*-\vec{p}^* e^TG^*\vec{p}^*,
\label{ODE*}
\end{equation}
\end{linenomath*}
where $e=(1,1,1)^T$ and $G^*$ in Eq. (\ref{Matrixx}). If we let the sum of each column of $G^*$ is the same, \emph{i.e.}
\begin{linenomath*}
$$\alpha_S=\alpha_B=\alpha_L=\kappa,$$
\end{linenomath*}
then Eq. (\ref{ODE*})
becomes
\begin{linenomath*}
\begin{equation}
\frac{d \vec{p}^*}{d t}=(G^*-\kappa I)\vec{p}^*,
\label{ODE3}
\end{equation}
\end{linenomath*}
where $I$ is identity matrix.
If we denote $H=(G^*-\kappa I)^T$, it can be shown that $H$ satisfies the
conditions (\ref{Q1}) and (\ref{Q2}) for the $Q$-matrix (see \ref{appendix2}).
In other words, the Kolmogorov forward equation Eq. (\ref{MC}) is a
special linear case of the nonlinear proportion equation Eq. (\ref{ODE*}).
This relation implies that, when the division rates of the three phenotypes are the same,
the dynamics of the phenotypic proportion can equivalently be captured
by the Markov chain model where only the phenotypic transitions are accounted for.
Otherwise, the Markov chain model may oversimplify the phenotypic dynamics with unequal division rates.
Interestingly, it was reported in Gupta \emph{et al}'s experiment that the subpopulations of S, B and
L phenotypes have the same ``doubling time'' \cite{gupta2011stochastic}, which justified
their application of the Markov chain model.
However, as mentioned in the end of Sec. \ref{section3.1}, Eq. (\ref{ODExx}) is
valid only in relatively short time scale. For larger time scales,
it is unreasonable to model the three-phenotypic dynamics by the Markov chain model
taking no account of different capabilities of divisions by cancer stem cells (unlimited) and
non-stem cancer cells (limited), even if they have the same division rate.
Therefore, one should be cautious about the application of the Markov chain in modeling
cell-state dynamics.

\subsection{Path-wise convergence of the MPB model}
\label{sectionIII}

We have seen that the MPB model provides a unified framework for the ODEs model and Markov chain model.
In this section, we will show path-wise convergence of the MPB model, which provides a much stronger concept of
stability by which both the stable steady-state behavior of the ODEs model and the equilibrium distribution of the Markov chain model
will serve as average-level stabilities of the MPB model.

Much attention has long been paid to the limit theorems of multi-type branching processes by mathematicians
\cite{athreya1968some,kesten1967limit,janson2004functional,yakovlev2010limiting}.
Here we are not going to discuss the rigorous mathematical theory in general
(which is the focus of our another work \cite{jiang2014cell}). Instead
we are more interested in the specific results related to the phenotypic equilibrium, \emph{i.e.}
the conditions under which $\vec{p}$ converges to a positive vector $\vec{\mu}$.
Unlike the $\vec{p}$ in Theorem \ref{Thm1}, the $\vec{p}$ here is stochastic.
The ``convergence'' here means \emph{almost sure convergence}.
That is, if the convergence of $\vec{p}$ holds, almost all the stochastic paths
will tend to a fixed equilibrium (also termed \emph{path-wise convergence}).

We present our main results in the following two theorems (see \ref{appendix3} for the proofs and mathematical details):

\begin{theorem}
If $G$ in Eq. (\ref{Matrix2}) is irreducible and its Perron-Frobenius eigenvalue is positive,
then $\vec{p}$ will tend to a fixed positive vector $\vec{\mu}$ almost surely as $t\rightarrow\infty$
conditioned on non-extinction of the population.
\label{Thm2}
\end{theorem}

\begin{theorem}
Assume that \\
(1) all the phenotypic transition rates are zero, i.e. $\beta_B$, $\beta_L$ and
$\gamma_B$, $\gamma_L$ are zero;\\
(2) $\alpha_S>0$, $\alpha_B>0$ and $\alpha_L>0$;\\
(3) $P_i>0$ ($1\leq i\leq 5$) and $P_1>P_2+P_3$;\\
then $\vec{p}$ will tend to a fixed positive vector $\vec{\mu}$ almost surely as $t\rightarrow\infty$
conditioned on non-extinction of stem like cells.
\label{Thm3}
\end{theorem}

The above two theorems are applicable to different cases. Theorem \ref{Thm2} corresponds to the case with phenotypic plasticity,
since the irreducibility of $G$ is satisfied as long as the conversions
between different phenotypes can happen. In contrast, Theorem \ref{Thm3} corresponds to the case without
phenotypic plasticity, since all the phenotypic transition rates are assumed to be zero.
Interestingly, even though the assumptions of the two theorems are basically different,
both of them can lead to the path-wise convergence $\vec{p}$. Furthermore, it is easy to see that the path-wise convergence of
$\vec{p}^*$ is implied by Theorems \ref{Thm2} and \ref{Thm3}:
\begin{corollary}
$\vec{p}^*$ will tend to a fixed positive vector $\vec{\mu}^*$ almost surely as $t\rightarrow\infty$
under the conditions in either Theorem \ref{Thm2} or Theorem \ref{Thm3}.
\label{cor2}
\end{corollary}
Figs. 2 and 3 illustrate the path-wise convergence of $\vec{p}^*$ implied by Theorems \ref{Thm2} and \ref{Thm3}
respectively by using stochastic simulations (\ref{appendix4} shows the simulations for $\vec{p}$ in details).
In both cases, even though all the stochastic paths fluctuate at the beginning of the process,
the proportions of S, B and L cells eventually converge to their equilibrium proportions as time passes.
Since the path-wise convergence indicates the stability of (almost) every stochastic sample,
the convergence of the mean dynamics just follows from it by averaging all the stochastic samples
(see lower panels of Figs. 2 and 3). Note that both the Kolmogorov forward equation of the Markov chain and the ODEs model
can be seen as the mean dynamics of the phenotypic proportions; their stabilities just correspond to the
average-level stabilities of the MPB model, which can be seen as direct results of the path-wise convergence.
In this way, the path-wise convergence provides a deeper understanding to the phenotypic equilibrium
from the stochastic point of view.

\begin{figure}
\begin{center}
\includegraphics[width=\textwidth]{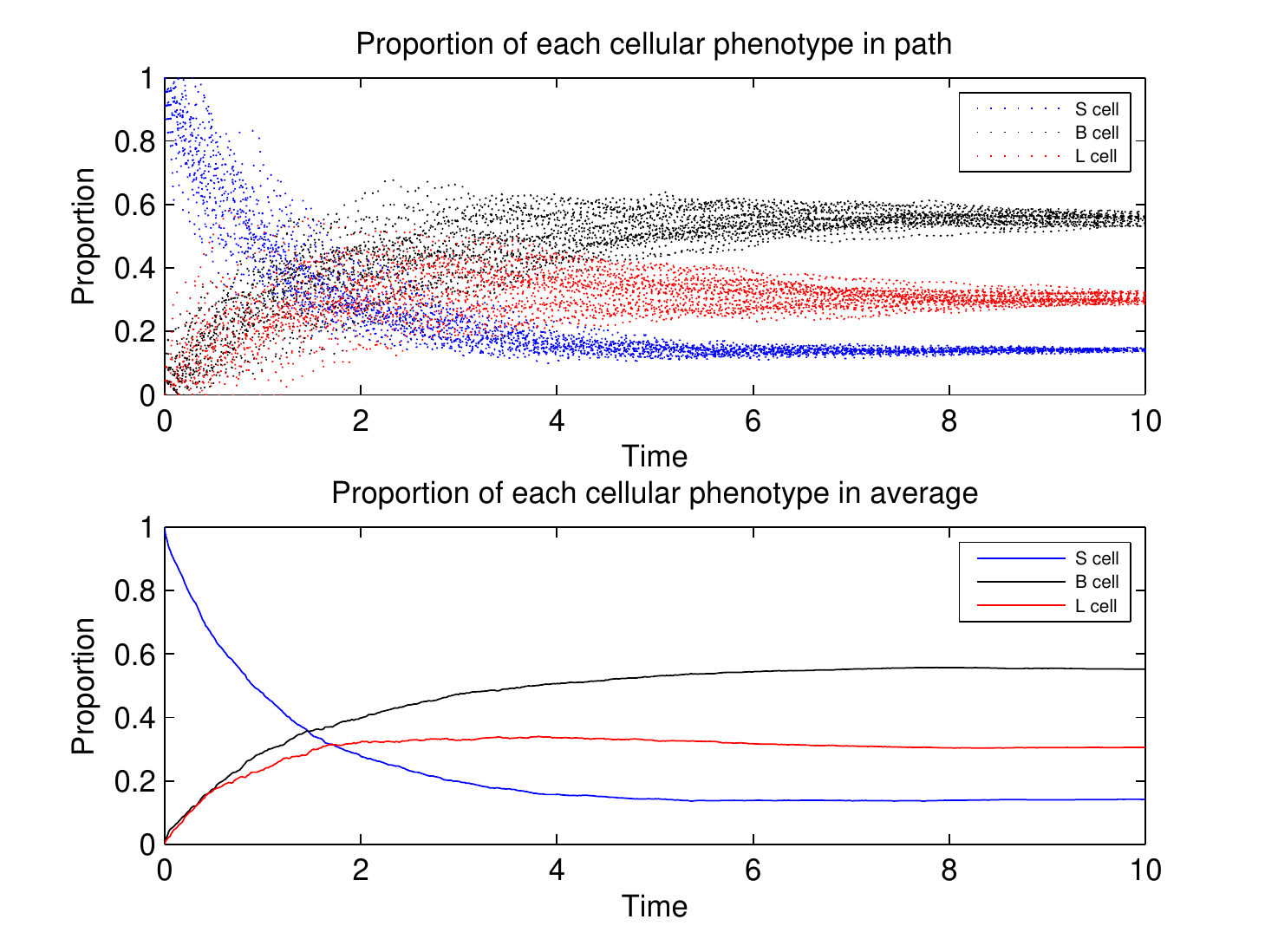}
\caption{Stochastic simulations for the case with phenotypic plasticity (Theorem \ref{Thm2}). Upper panel shows the stochastic path-wise dynamics of the phenotypic proportions of S (blue), B (black) and L (red). The initial numbers of S, B and L cells are assumed to be 20, 0 and 0 respectively, that is, the initial proportions of S, B and L cells are 100\%, 0\% and 0\%.
According to the assumptions in Theorem \ref{Thm2}, we set $m=10$;
$\alpha_S=0.8$, $P_1=0.3$, $P_2=0.2$, $P_3=0.2$, $P_4=0.15$, $P_5=0.15$;
$\alpha_B=0.6$, $\alpha_{B_m}=0.3$, $\beta_B=0.1$, $\gamma_B=0.05$;
$\alpha_L=0.7$, $\alpha_{L_m}=0.3$, $\beta_L=0.13$, $\gamma_L=0.2$.
Thirty stochastic samples for each phenotype were produced. It is shown that even though the stochastic paths fluctuate at the beginning of the process, the proportions of S, B and L phenotypes eventually path-wisely tend to their equilibrium proportions respectively. Lower panel shows the mean dynamics of the phenotypic proportions by averaging all the thirty samples shown in upper panel.}
\end{center}
\end{figure}

\begin{figure}
\begin{center}
\includegraphics[width=\textwidth]{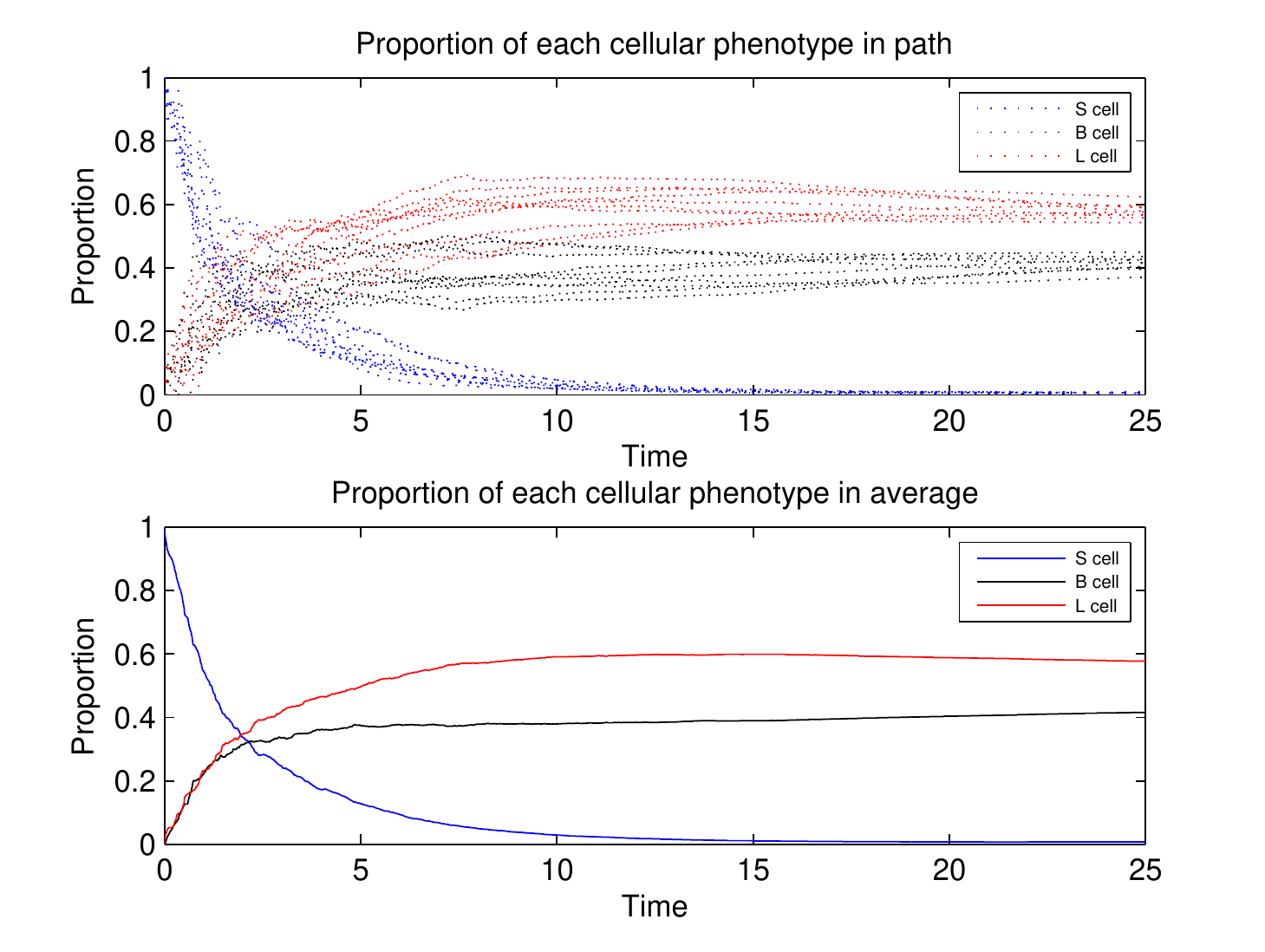}
\caption{Stochastic simulations for the case without phenotypic plasticity (Theorem \ref{Thm3}).
The initial cell numbers of S, B and L cells are also assumed to be 20, 0 and 0 respectively.
According to the assumptions in Theorem \ref{Thm3}, we set $m=10$;
$\alpha_S=0.8$, $P_1=0.5$, $P_2=0.14$, $P_3=0.16$, $P_4=0.1$, $P_5=0.1$;
$\alpha_B=0.4$, $\alpha_{B_m}=0.3$, $\beta_B=0$, $\gamma_B=0$;
$\alpha_L=0.45$, $\alpha_{L_m}=0.3$, $\beta_L=0$, $\gamma_L=0$.
Ten stochastic samples for each phenotype were produced. Upper panel shows the path-wise convergence of the phenotypic proportions. Lower panel shows the average-level stability of the mean dynamics.}
\end{center}
\end{figure}

As the end of this section, it is noteworthy to emphasize that, according to Theorem \ref{Thm3}, the
phenotypic equilibrium can still happen in the paradigm of conventional cancer stem cell theory.
The assumptions in Theorem \ref{Thm3} together indicate the cellular hierarchy proposed by the cancer stem cell
theory \cite{dalerba2007cancer}. That is, cancer stem cells (S cells)
are capable of differentiation into other more committed non-stem cancer cells (B and L cells)
but not vice versa. In this way, cancer stem cells are at the apex of this cellular hierarchy.
Moreover, the assumption ``$P_1>P_2+P_3$'' implies the \emph{dominance} of S phenotype during the growth of the population.
To show this, note that $\alpha_S\left(P_1-P_2-P_3\right)$ is the eigenvalue corresponding to S phenotype of $G$ in
Eq. (\ref{matrix3}), which is the only positive eigenvalue of $G$ provided ``$P_1>P_2+P_3$''
(see \ref{appendix3}). In other words, instead of the phenotypic plasticity, Theorem \ref{Thm3} also
gives an alternative explanation to the phenotypic equilibrium in the framework of the cancer stem cell theory,
as long as the cancer stem cell phenotype is dominant in the population. However, it is interesting to see that
the convergence rate of the case in Fig. 2 is faster than that of the case in Fig. 3, even though they both
give rise to the path-wise convergence. This suggests that perhaps the convergence rate (rather than the convergence itself)
could serve as an indicator to distinguish the models with and without phenotypic plasticity, which might be another
meaningful research topic in future.

\section{Conclusions}

In this study, we have presented a multi-phenotype branching model of cancer cells.
On one hand, this model can serve as an underlying model from which the ODEs model and
the Markov chain model can be deduced. On the other hand, the almost sure convergence of the
model enhances our understanding of the phenotypic equilibrium, from average-level stability
to path-wise convergence. Furthermore, our results have indicated that, even though the
phenotypic plasticity facilitates the phenotypic equilibrium, it is not indispensable in some cases.
It has been shown that the conventional cancer stem cell model can also stabilize the mixture of
the phenotypic proportions, providing an alternative explanation to the phenotypic equilibrium.

Moreover, it should be noted that even though this work is focused on
the issue of cancer, our methods can conveniently be used to
more generalized cell population dynamics \cite{jiang2014cell}.
To further reveal the biological mechanisms of the phenotypic equilibrium,
more detailed dynamic models of cancer cells are needed. For instance,
the hypothesis of cooperation among cancer cells has been put forward
\cite{axelrod2006evolution}. In particular, self-sufficiency of certain growth signals
of cancer cells supports the concept of mutualism and could be an important mechanism
supporting the phenotypic equilibrium. Therefore, the models of capturing the
interactions among cancer cells, \emph{e.g.} evolutionary game models \cite{nowak2004evolutionary},
could be a promising research direction in future.
Furthermore, the genetic and epigenetic state networks \cite{huang2012molecular,wang2013phage} of cancer  will enable us to explore the
molecular mechanisms of the phenotypic equilibrium, which are poorly understood.
The network methods have successfully been used to investigate the processes of cellular pluripotent reprogramming \cite{wang2012global}
and epithelial-mesenchymal transitions (EMT) \cite{jolly2014towards}.
Note that EMT could play a key role in regulating the phenotypic heterogeneity in cancer \cite{may2011epithelial},
further studies on it should be another important tasks in future plans.

\section*{Acknowledgements}
D. Z. acknowledges the generous sponsorship from
the National Natural Sciences Foundation of China (No. 11401499),
the Natural Science Foundation of Fujian Province of China (No. 2015J05016),
and the Fundamental Research Funds for the Central Universities (No. 20720140524).
Y. N. is supported by National Natural Science Foundation of China (No.11401594) and the New Teachers' Specialised
Research Fund for the Doctoral Program from Ministry of Education of China (No.20120162120096).

\appendix
\section{Expanded form of Eq. (\ref{CME})}
\label{appendix1}

Here we show more details about the master equation Eq. (\ref{CME})
\begin{linenomath*}
\begin{equation*}
\frac{d\textrm{Pr}(\vec{x};t)}{dt}=\sum_{\vec{x}'\neq \vec{x}}T_{\vec{x}'\rightarrow \vec{x}}\textrm{Pr}(\vec{x}';t)-\sum_{\vec{x}'\neq \vec{x}}T_{\vec{x}\rightarrow \vec{x}'}\textrm{Pr}(\vec{x};t).
\end{equation*}
\end{linenomath*}
To obtain the expanded form of  Eq. (\ref{CME}), we need to confirm all possible $T_{\vec{x}'\rightarrow \vec{x}}$ and $T_{\vec{x}'\rightarrow \vec{x}}$.
Based on the model assumptions, we can calculate $T_{\vec{x}'\rightarrow \vec{x}}$ and $T_{\vec{x}'\rightarrow \vec{x}}$
correspondingly. For example, let $\vec{x}'=(x_1-1, x^{(0)}_2,...,x^{(m)}_2, x^{(0)}_3,...,x^{(m)}_3)^T$,
the event ``$(x_1-1, x^{(0)}_2,...,x^{(m)}_2, x^{(0)}_3,...,x^{(m)}_3)^T\rightarrow (x_1, x^{(0)}_2,...,x^{(m)}_2, x^{(0)}_3,...,x^{(m)}_3)^T$'' will happen if any one of
``$S\overset{\alpha_{S}P_1}{\longrightarrow} S+S$'' happens. Since the number of S cells is $x_1-1$ in current population,
the transition rate should be $(x_1-1)\times \alpha_{S}P_1$. On the other hand,
the reaction ``$S\overset{\alpha_{S}P_1}{\longrightarrow} S+S$'' can also lead to the transition from
$(x_1, x^{(0)}_2,...,x^{(m)}_2, x^{(0)}_3,...,x^{(m)}_3)^T\rightarrow (x_1+1, x^{(0)}_2,...,x^{(m)}_2, x^{(0)}_3,...,x^{(m)}_3)^T$ with rate $x_1\times \alpha_S P_1$.
Along this way we can determine all the transition rates similarly. Therefore,
\begin{linenomath*}
\begin{eqnarray*}
&&\sum_{\vec{x}'\neq \vec{x}}T_{\vec{x}'\rightarrow \vec{x}}\textrm{Pr}(\vec{x}';t)=\\
&& (x_1-1)\alpha_{S}P_1\textrm{Pr}(x_1-1, x^{(0)}_2,...,x^{(m)}_2, x^{(0)}_3,...,x^{(m)}_3;t)\\
&&+(x_1+1)\alpha_S P_2\textrm{Pr}(x_1+1, x^{(0)}_2-2,...,x^{(m)}_2, x^{(0)}_3,...,x^{(m)}_3;t)\\
&&+(x_1+1)\alpha_S P_3\textrm{Pr}(x_1+1, x^{(0)}_2,...,x^{(m)}_2, x^{(0)}_3-2,...,x^{(m)}_3;t)\\
&&+x_1\alpha_S P_4\textrm{Pr}(x_1, x^{(0)}-1,...,x^{(m)}_2, x^{(0)}_3,...,x^{(m)}_3;t)\\
&&+x_1\alpha_S P_3\textrm{Pr}(x_1, x^{(0)}_2,...,x^{(m)}_2, x^{(0)}_3-1,...,x^{(m)}_3;t)\\
&&+\sum_{i=1}^{m}(x^{(i-1)}_2+1)\alpha_B\textrm{Pr}(x_1,..., x^{(i-1)}_2+1,x^{(i)}_2-2,...;t)\\
&&+(x^{(m)}_2+1)\alpha_{B_m}\textrm{Pr}(x_1, x^{(0)}_2,...,x^{(m)}_2+1, x^{(0)}_3,...,x^{(m)}_3;t)\\
&&+\sum_{i=0}^{m}(x^{(i)}_2+1)\beta_B\textrm{Pr}(x_1-1,..., x^{(i)}_2+1,...;t)\\
&&+\sum_{i=0}^{m}(x^{(i)}_2+1)\gamma_B\textrm{Pr}(x_1,..., x^{(i)}_2+1,...,x^{(i)}_3-1,...;t)\\
&&+\sum_{i=1}^{m}(x^{(i-1)}_3+1)\alpha_L\textrm{Pr}(x_1,..., x^{(i-1)}_3+1,x^{(i)}_3-2,...;t)\\
&&+(x^{(m)}_3+1)\alpha_{L_m}\textrm{Pr}(x_1, x^{(0)}_2,...,x^{(m)}_2, x^{(0)}_3,...,x^{(m)}_3+1;t)\\
&&+\sum_{i=0}^{m}(x^{(i)}_3+1)\beta_L\textrm{Pr}(x_1-1,..., x^{(i)}_3+1,...;t)\\
&&+\sum_{i=0}^{m}(x^{(i)}_3+1)\gamma_L\textrm{Pr}(x_1,..., x^{(i)}_2-1,...,x^{(i)}_3+1,...;t),\\
\end{eqnarray*}
\end{linenomath*}
and
\begin{linenomath*}
\begin{eqnarray*}
&&\sum_{\vec{x}'\neq \vec{x}}T_{\vec{x}\rightarrow \vec{x}'}\textrm{Pr}(\vec{x};t)=x_1\sum_{i=1}^5\alpha_S P_i\textrm{Pr}(\vec{x};t)\\
&&+\sum_{i=0}^{m-1}\alpha_B x^{(i)}_2\textrm{Pr}(\vec{x};t)
+\alpha_{B_m}x^{(m)}_2\textrm{Pr}(\vec{x};t)+\sum_{i=0}^{m}(\beta_B+\gamma_B) x^{(i)}_2\textrm{Pr}(\vec{x};t)\\
&&+\sum_{i=0}^{m-1}\alpha_L x^{(i)}_3\textrm{Pr}(\vec{x};t)
+\alpha_{L_m}x^{(m)}_3\textrm{Pr}(\vec{x};t)+\sum_{i=0}^{m}(\beta_L+\gamma_L) x^{(i)}_3\textrm{Pr}(\vec{x};t).
\end{eqnarray*}
\end{linenomath*}

\section{Proportion equation and Kolmogorov forward equation}
\label{appendix2}

Firstly we show how to derive the proportion equation (\ref{ODE2}) from the number equation (\ref{ODE1}).
Let us rewrite the matrix form of Eq. (\ref{ODE1}) into component form
\footnote{The dimension of $G$ is $2m+3$, for simplicity we let $2m+3=n$.}:
\begin{linenomath*}
$$\frac{d\langle X_i\rangle}{dt}=g_{i1}\langle X_1\rangle+g_{i1}\langle X_2\rangle+...+g_{in}\langle X_n\rangle.$$
\end{linenomath*}
Note that
\begin{linenomath*}
$$p_i=\frac{\langle X_i \rangle}{\langle X_1+X_2+...+X_n \rangle}=\frac{\langle X_i \rangle}{N},$$
$$\frac{d\langle X_i\rangle}{dt}=\frac{d(p_i N)}{dt}=p_i\frac{dN}{dt}+N\frac{dp_i}{dt},$$
\end{linenomath*}
then
\begin{linenomath*}
\begin{align*}
\frac{dp_i}{dt}
&=\frac{1}{N}\frac{d\langle X_i\rangle}{dt}-\frac{p_i}{N}\frac{dN}{dt}\\
&=\sum_{j=1}^{n}(g_{ij}p_j)-p_i\sum_{k=1}^{n}\sum_{j=1}^{n}(g_{kj}p_j).
\end{align*}
\end{linenomath*}
When turning the above component form back to the matrix form, we get Eq. (\ref{ODE2})
\begin{linenomath*}
\begin{equation*}
\frac{d \vec{p}}{d t}=G\vec{p}-\vec{p} e^TG\vec{p}.
\end{equation*}
\end{linenomath*}
Similarly, we can also get the proportion equation Eq. (\ref{ODEx}) for $\vec{p}^*$
\begin{linenomath*}
\begin{equation}
\frac{d \vec{p}^*}{d t}=G^*\vec{p}^*-\vec{p}^* e^TG^*\vec{p}^*.
\label{ODE4}
\end{equation}
\end{linenomath*}
In what follows we show how the proportion equation Eq. (\ref{ODE4})
relates to the Kolmogorov forward equation of continuous-time Markov chain.
If the column sums of $G^*$ are the same and equal to $\kappa$, \emph{i.e.}
\begin{linenomath*}
$$\alpha_{S}=\alpha_{B}=\alpha_{L}=\kappa,$$
\end{linenomath*}
then
\begin{linenomath*}
\begin{eqnarray*}
\frac{d \vec{p}^*}{d t}=G^*\vec{p}^*-\kappa\vec{p}^*=H^T\vec{p}^*.
\end{eqnarray*}
\end{linenomath*}
where $H=(G^*-\kappa I)^T$ and $I$ is identity matrix. For any $i\neq j$, it is easy to see that $h_{ij}=g^*_{ji}$.
Note that all the off-diagonal elements
of $G^*$ are non-negative, so $h_{ij}\geq0$, satisfying the condition Eq. (\ref{Q1}).
Meanwhile, $h_{ii}=g^*_{ii}-\kappa$. Note that $\kappa=\sum_{j}g^*_{ji}$,
\begin{linenomath*}
$$h_{ii}=g^*_{ii}-\kappa=-\sum_{j\neq i}g^*_{ji}=-\sum_{j\neq i}h_{ij},$$
\end{linenomath*}
satisfying the condition Eq. (\ref{Q2}). Hence $H$ corresponds to
the $Q$-matrix of a continuous-time Markov chain.

\section{Proof of Theorem \ref{Thm1}}
\label{appendix2+}

First of all, we have two remarks on $G$ in Eq. (\ref{Matrix2}):
\begin{itemize}
  \item Note that the off-diagonal elements of $G$ are all non-negative, we call $G$ an ML-matrix (see Chapter 2 in \cite{seneta1981non}).
  For sufficiently large $\tau$, $G+\tau I$ is a non-negative matrix (I is an identity matrix). In other words, ML-matrix is \emph{essentially
  non-negative}. The ML-matrix $G$ is said to be irreducible if $G+\tau I$ is irreducible
  \footnote{A non-negative matrix $M$ is said to be irreducible, if for every pair of indices $i$ and $j$, there exists a natural number $k$ such that $[M^k]_{ij}$ is larger than 0.}.
  \item When $G$ is irreducible, from Theorem 2.6 in \cite{seneta1981non}, there exists a Perron-Frobenius eigenvalue $\lambda_1$ satisfying that \emph{1)}
  $\lambda_1$ is real and $\lambda_1>\textrm{Re}\lambda$ for any eigenvalue $\lambda\neq \lambda_1$; \emph{2)} $\lambda_1$ is simple, \emph{i.e.}
  a simple root of the characteristic equation of $G$; \emph{3)}  $\lambda_1$ is associated with (up to constant multiples) unique positive right eigenvector $\vec{\mu}$. Here we assume that
  $\vec{\mu}$ is the normalized right eigenvector of $\lambda_1$, that is, $\mu_1+\mu_2+...+\mu_n=1$.
\end{itemize}

Since $\lambda_1$ is simple, the solution of Eq. (\ref{ODE1}) can be expressed as
\begin{linenomath*}
\begin{equation}
\begin{split}
&\langle \vec{X} \rangle=c_{1,1}\vec{\mu}e^{\lambda_{1}t}+\sum_{j=2}^{m}\sum_{l=1}^{m_j}c_{j,l}\sum_{i=1}^{m_j}\vec{r^{j}_{l,i}}t^{i-1}e^{\lambda_{j}t},
\end{split}
\label{solution}
\end{equation}
\end{linenomath*}
where $\lambda_{1},\lambda_{2},\cdots{}\lambda_{m}$ are the different eigenvalues of $G$,
$m_j$ is the algebraic multiplicity of $\lambda_{j}$,
$\vec{r^j_{l,i}}$ is the corresponding eigenvector of $\lambda_{j}$,
$c_{j,l}$ is determined by initial states. Suppose $c_{1,1}\neq{}0$, since Re$\lambda_i<\lambda_1~(i\neq 1)$,
\begin{linenomath*}
\begin{equation*}
\begin{split}
&\lim_{t\rightarrow +\infty}\frac{\langle \vec{X} \rangle}{c_{1,1}e^{\lambda_{1}t}}=\vec{\mu}+\lim_{t\rightarrow +\infty}\sum_{j=2}^{m}\sum_{l=1}^{m_j}
\frac{c_{j,l}}{c_{1,1}}\sum_{i=1}^{m_j}\vec{r^{j}_{l,i}}t^{i-1}e^{(\lambda_{j}-\lambda_1)t}= \vec{\mu}.
\end{split}
\end{equation*}
\end{linenomath*}
Thus
\begin{linenomath*}
$$\vec{p}=\frac{\langle \vec{X} \rangle}{\langle X_1+X_2+...+X_n \rangle}=\frac{\langle \vec{X} \rangle/c_{1,1}e^{\lambda_{1}t}}{\langle X_1+...+X_n \rangle/c_{1,1}e^{\lambda_{1}t}}
\rightarrow \frac{\vec{\mu}}{\mu_1+...+\mu_n}=\vec{\mu}>0.$$
\end{linenomath*}

Before completing the proof, we need to discuss the case $c_{1,1}=0$. In this case, the above argument
does not work. However, since fluctuations are inevitable in real world,
$c_{1,1}=0$ will hardly happen in reality. To show this, let $t=0$ in Eq. (\ref{solution})
\begin{linenomath*}
\begin{equation*}
\begin{split}
&c_{1,1}\vec{\mu}+\sum_{j=2}^{m}\sum_{l=1}^{m_j}c_{j,l}\vec{r^{j}_{l,1}}=\langle\vec{X_0}\rangle.
\end{split}
\end{equation*}
\end{linenomath*}
This is a linear equation of $c_{j,l}$. By Cramer's Rule we have
\begin{linenomath*}
\begin{equation*}
\begin{split}
&c_{1,1}=\frac{\textrm{det}|B^*|}{\textrm{det}|B|},
\end{split}
\end{equation*}
\end{linenomath*}
where $B=[\vec{\mu}\ \vec{r^{2}_{1,1}}\ \vec{r^{2}_{2,1}}\cdots \vec{r^{m}_{m_m,1}}]$, $B^*$ is just $B$ with
its first column replaced by $\langle\vec{X_0}\rangle$.
It is easy to add a small perturbation $\varepsilon{}\vec{v}$ to $\langle\vec{X_0}\rangle$,
so that all the columns of $B^*$ are linear independent, hence $c_{1,1}\neq{}0$ holds.

\section{Proofs of Theorems \ref{Thm2} and \ref{Thm3}}
\label{appendix3}

The proofs of Theorems \ref{Thm2} and \ref{Thm3} are both on the basis of the following lemma
\footnote{As far as we know, Theorem 5 in \cite{jiang2014cell} requires minimal constraint to the path-wise
convergence of our concern. However, it should be noted that technically our main results
can also be proved based on Theorem 3.1 in \cite{janson2004functional}.}:
\newtheorem{lemma}{Lemma}
\begin{lemma}
[Theorem 5 in \cite{jiang2014cell}] Assume that the Perron-Frobenius eigenvalue
$\lambda_1$ of $G$ in Eq. (\ref{Matrix2}) is simple and positive.
Conditioned on essential non-extinction, $\vec{p}$ will tend to $\vec{\mu}$ almost surely as $t\rightarrow\infty$.
$\vec{\mu}$ is the normalized right eigenvector of $\lambda_1$, which is non-negative.
\label{lemma}
\end{lemma}

For proving Theorems \ref{Thm2} and \ref{Thm3}, firstly we need to explain the concept of essential non-extinction.
We are not going to discuss the general mathematical definition of it (see Sec. 4.2 in \cite{jiang2014cell}).
In our MPB model, essential non-extinction specifically means non-extinction of the phenotype
corresponding to the Perron-Frobenius eigenvalue $\lambda_1$.
For Theorem \ref{Thm3}, the assumptions (2) and (3) implies that the Perron-Frobenius eigenvalue
of $G$ is $g_{11}=\alpha_S(P_1-P_2-P_3)$ (we will show this later). In other words, the essential non-extinction here just
means non-extinction of stem-like phenotype. For Theorem \ref{Thm2}, since $G$ is irreducible, it is possible
for any two phenotypes to (directly or indirectly) inter-convert into each other. In this case, non-extinction of one particular phenotype is
equivalent to non-extinction of any phenotype. This implies that, no matter which phenotype corresponds to $\lambda_1$,
to guarantee essential non-extinction, it is sufficient to assume non-extinction of the population in general.
Therefore, the conditions provided in Theorems \ref{Thm2} or \ref{Thm3} ensure the essential non-extinction of the model.

We now start to prove the two theorems. On one hand, we need to show that $\lambda_1$ of $G$ is simple and positive in both theorems.
On the other hand, since Lemma \ref{lemma} only concludes the non-negativity of $\vec{\mu}$,
we need to further show the positivity of $\vec{\mu}$.
The proof for Theorem \ref{Thm2} is straightforward, since we assume that $G$ is irreducible,
according to the second remark in \ref{appendix2+}, $\lambda_1$ is simple and $\mu$ is positive.
Note that $\lambda_1$ is also assumed positive, by Lemma \ref{lemma} we have $\vec{p}\rightarrow\vec{\mu}>0$ almost surely as $t\rightarrow\infty$.

For Theorem \ref{Thm3}, according to the assumptions, $G$ reduces to a lower triangular matrix as follows
\begin{linenomath*}
\begin{equation}
G=[g_{ij}]=\left(\begin{smallmatrix}
\alpha_S\left(P_1-P_2-P_3\right) & 0 & \cdots & \cdots & \cdots & \cdots & 0 \\
\alpha_S\left(2P_2+P_4\right) & -\alpha_{B} & 0 & \cdots & \cdots & \cdots & 0  \\
0 & 2\alpha_{B} & -\alpha_{B} & 0 & \cdots & \cdots & 0 \\
\cdots & \cdots & \cdots & \cdots & \cdots & \cdots & \cdots\\
\alpha_S\left(2P_3+P_5\right) & & & 0 & -\alpha_L & \cdots & 0\\
0 & \cdots & \cdots & \cdots & 2\alpha_L & -\alpha_L & \cdots \\
\cdots & \cdots & \cdots & \cdots & \cdots & \cdots & \cdots\\
\end{smallmatrix}\right).
\label{matrix3}
\end{equation}
\end{linenomath*}
It is easy to know that the eigenvalues of $G$ correspond to the diagonal elements. By assumptions (2) and (3),
$\lambda_1=\alpha_S\left(P_1-P_2-P_3\right)$ is the Perron-Frobenious eigenvalue which is positive and simple.
By Lemma \ref{lemma}, we have $\vec{p}\rightarrow\vec{\mu}$ almost surely as $t\rightarrow\infty$,
where $\vec{\mu}$ is the normalized right eigenvector of $\lambda_1$. To complete the proof, we need to show
that $\vec{\mu}$ is positive. Note that $\vec{\mu}$ satisfies the following equation
\begin{linenomath*}
\begin{equation}
G\vec{\mu}=\lambda_1\vec{\mu}.
\end{equation}
\end{linenomath*}
By expanding this equation, we have
\begin{linenomath*}
\begin{equation}
\begin{cases}
\lambda_1\mu_1=\lambda_1\mu_1;\\
\alpha_S\left(2P_2+P_4\right)\mu_1=(\alpha_{B}+\lambda_1)\mu_2;\\
2\alpha_B\mu_2=(\alpha_{B}+\lambda_1)\mu_3;\\
\vdots\\
2\alpha_B\mu_{m}=(\alpha_{B_{m}}+\lambda_1)\mu_{m+1};\\
\alpha_S\left(2P_3+P_5\right)\mu_1=(\alpha_{L}+\lambda_1)\mu_(m+2);\\
2\alpha_L\mu_{m+2}=(\alpha_{L}+\lambda_1)\mu_{m+3};\\
\vdots\\
2\alpha_L\mu_{2m+2}=(\alpha_{L_{m}}+\lambda_1)\mu_{2m+3}.\\
\end{cases}
\end{equation}
\end{linenomath*}
Suppose $\mu_1>0$, then we have
\begin{linenomath*}
$$\mu_2=\frac{\alpha_S\left(2P_2+P_4\right)}{\alpha_{B}+\lambda_1}\mu_1>0$$
\end{linenomath*}
since $\alpha_S\left(2P_2+P_4\right)>0$ and $\alpha_{B}+\lambda_1>0$.
With the same logic, we can show the positivity of $\mu_i$
recursively, which completes the final proof.

\section{Stochastic simulations for Theorems \ref{Thm2} and \ref{Thm3}}
\label{appendix4}

\begin{figure}
\begin{center}
\subfigure[] {\includegraphics[width=0.7\textwidth]{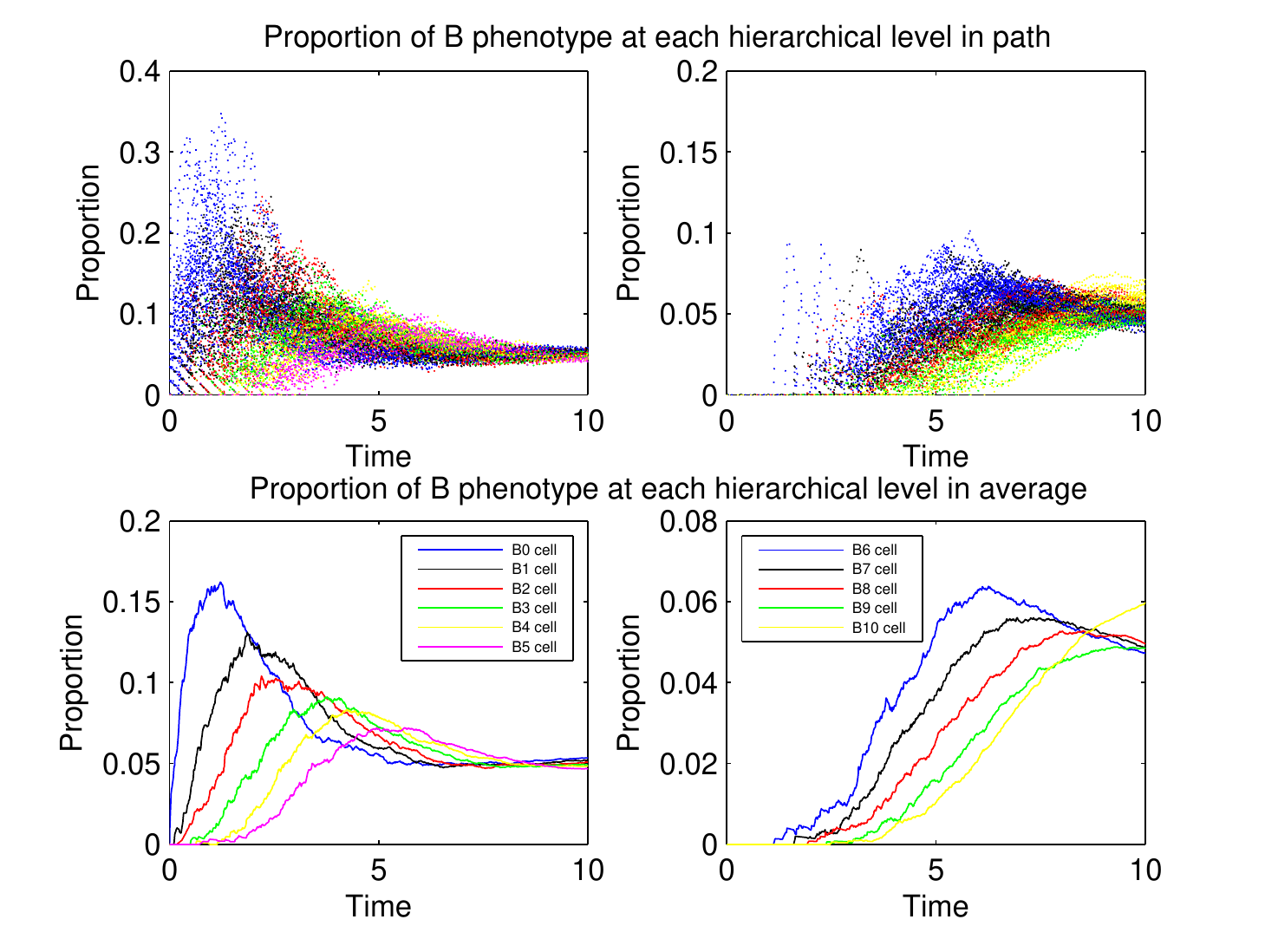}}
\subfigure[] {\includegraphics[width=0.7\textwidth]{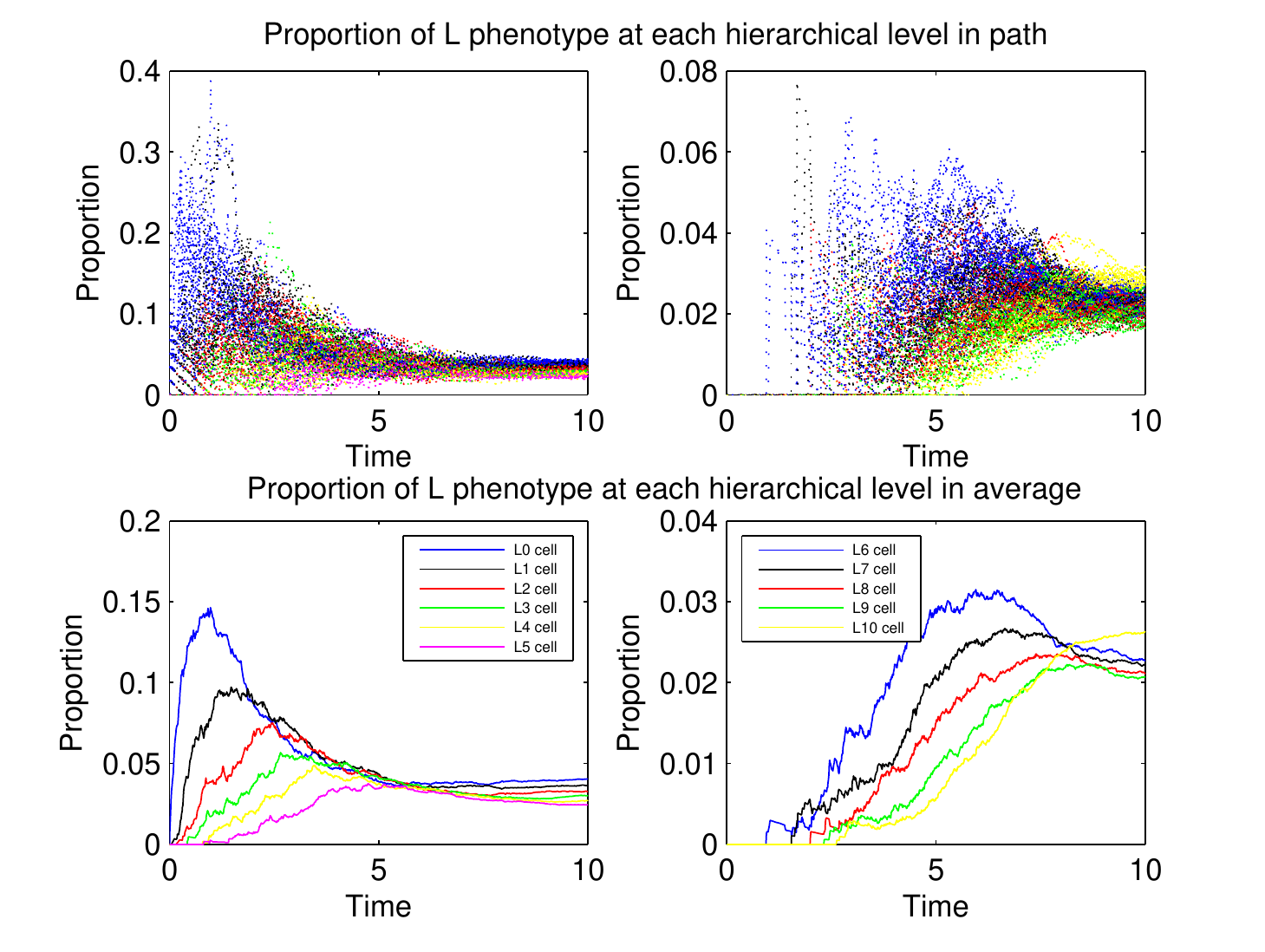}}
\caption{Illustration of Theorem \ref{Thm2}. The parameters are the same as those in Fig. 2. Dynamics of B and L phenotypes at each hierarchical level are shows in (a) and (b) respectively.
}
\label{FigE4}
\end{center}
\end{figure}

\begin{figure}
\begin{center}
\subfigure[] {\includegraphics[width=0.7\textwidth]{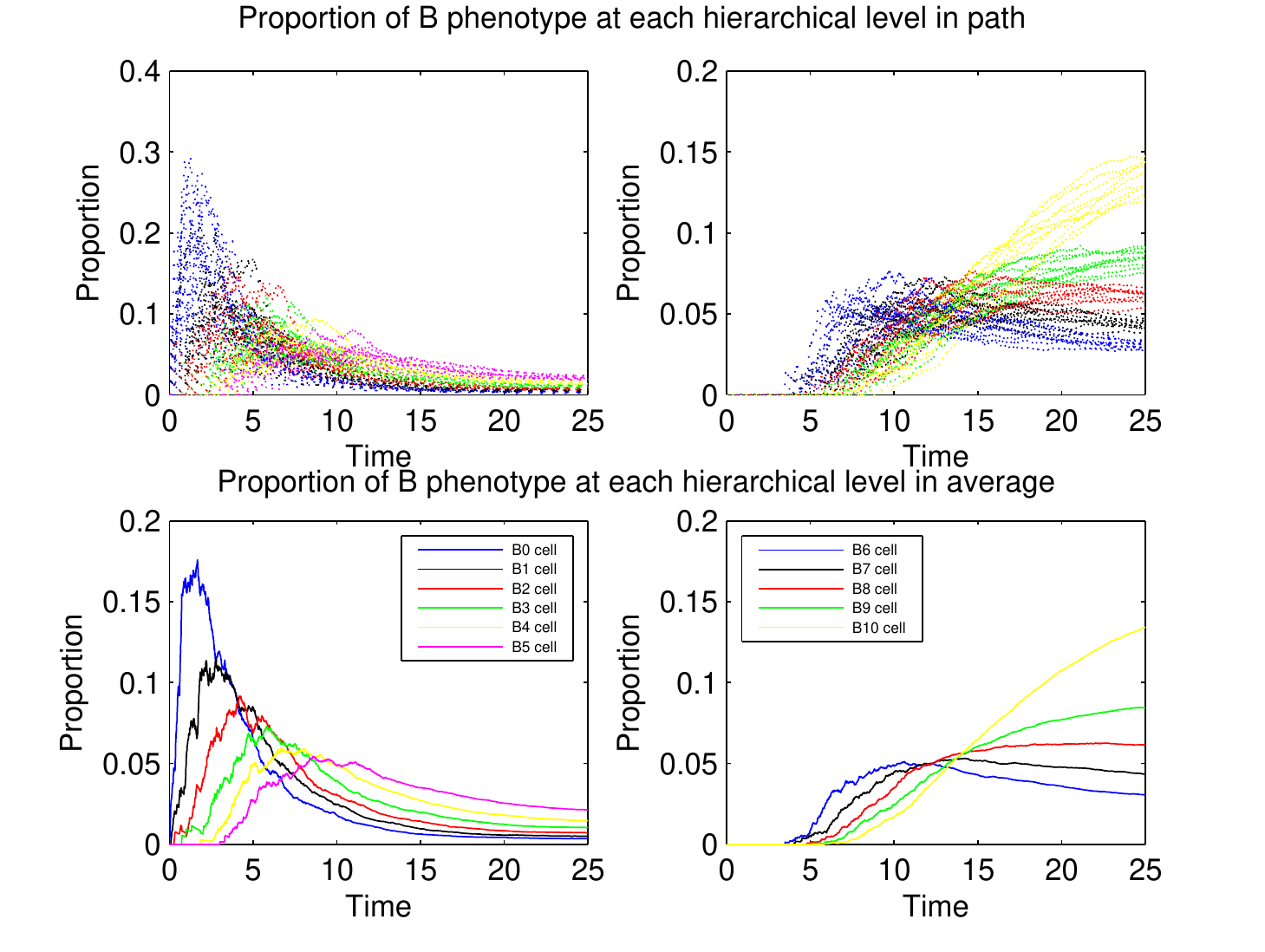}}
\subfigure[] {\includegraphics[width=0.7\textwidth]{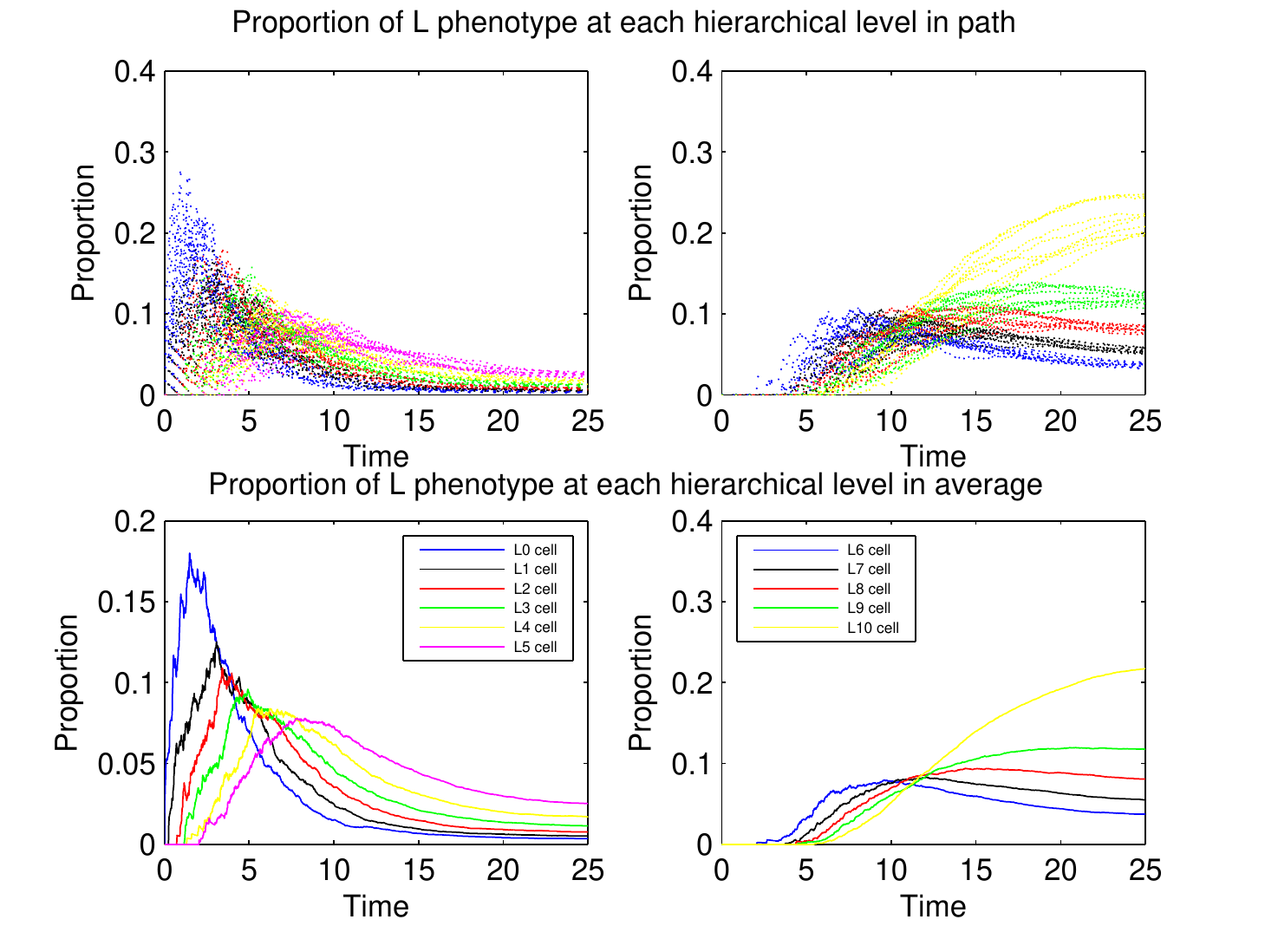}}
\caption{Illustration of Theorem \ref{Thm3}. The parameters are the same as those in Fig. 3.}
\label{FigE4}
\end{center}
\end{figure}

\end{document}